\documentstyle[twoside,fleqn,espcrc2,epsf]{article}
\include{tsumacro}

\newcommand{\AmS}{{\protect\the\textfont2
  A\kern-.1667em\lower.5ex\hbox{M}\kern-.125emS}}

\newcommand{\np}{Nucl. Phys. \underline}

\title{$\epsi$: theoretical results and updated phenomenological analysis}

\author{Laura Reina\address{Universit\'e Libre de Bruxelles, Service
de Physique Th\'eorique, \\ Boulevard du Triomphe, CP 225, B-1050
Brussels, Belgium}\thanks{Work done in collaboration with M. Ciuchini,
E. Franco and G. Martinelli of the University of Rome ``La
Sapienza''.}}
       
\begin{document}

\begin{abstract}
We present the results for the Next-to-Leading Order effective
hamiltonian for $\Delta S\!=\! 1$ decays, in presence of both QCD and
QED corrections, and update the existing theoretical predictions for
$\epsi$.
\end{abstract}

% typeset front matter (including abstract)
\maketitle

\section{Introduction}

During the last few years a big effort has been made in order to state
the theoretical predictions for K- and B-meson physics on a more solid
ground. Our interest in K- and B-physics is due both to theoretical
and experimental reasons. From the theoretical point of view, this
kind of physics is the natural framework to analyze and test the
mechanisms of flavour mixing and CP violation proposed in the Standard
Model and beyond. $\Delta F\!=\!2$ (F=flavour) mixings, $\Delta
F\!=\!1$ weak decays (and related asymmetries), rare K- and B-decays
provide in principle all the necessary tools for a full-fledged
analysis of this important aspect of the present
phenomenology of elementary particle physics. Several ``still
missing'' parameters of the Standard Model, tipically the ones related
to the top quark -- as the top mass ($m_t$) and couplings
($V_{td},\ldots$) -- can be constrained by CP violation and flavour
mixing results: a real alternative prediction to the pure electroweak
determinations, moreover improvable in the near future.

On the other hand, the nowadays theoretical enthusiasm is supported by
an exciting experimental scenario, where more and more precision is
reached in single measurements or expected in first generation
experiments.  This is expecially the case of B-physics, so extensively
discussed at this conference, and of $\epsi$, the {\it direct} CP
violation parameter in Kaon decays into two pions.  The smallness of
this CP violation parameter, due both to its {\it radiative} origin
and to its intrinsic dependence on the $\Delta I\!=\!1/2$ {\it rule}
\cite{peccei2}, is at the same time the origin of its interest and of
its illness.  $\epsi$ is today still compatible with zero, both
theoretically and experimentally. At the same time it is crucial to
state if it vanishes, in order to trust the Standard Model or
not. Thus, it is a real challenge to reduce more and more all the
uncertainties in the problem.

Our contribution in this direction has been the calculation of the
effective hamiltonian for $\Delta S\!=\!1$ weak decays at the
Next-to-Leading Order (NLO) in QCD and QED. The effective hamiltonian
formalism is the typical framework for the analytical calculation of
the physical amplitudes for weak processes, i.e. in a theory where
very large masses and mass gaps are present. Within this formalism,
the hamiltonian which describes a given physical process is expressed
as a linear combination of effective operators with certain Wilson
coefficients. In so doing, the perturbative and non-perturbative
realms factorize and we can fully exploit our analytical capabilities
in calculating the Wilson coefficients at a given perturbative
order, while taking the operator matrix elements from some
non-perturbative results.

They are precisely the non-perturbative inputs which represent the
``dark side'' of the problem, being affected by very large
uncertainties, which show up not only in the operator matrix elements,
but also in some physical input parameters (like the CKM parameters,
etc.) whose prediction relies on some ``long-distance physics''.

Taking ``long-distance physics'' from Lattice QCD (just a method among
others, to be predictive from a quantitative point of view), our
effort has been to state the perturbative calculation of the Wilson
coefficients on a more solid ground. A NLO effective hamiltonian
constitutes a more reliable result in many respects: the
$\mu$-dependence is greatly reduced ($\mu$ being the scale of matching
between Wilson coefficients and matrix elements); $\Lambda_{QCD}$ can
be properly taken from other experiments which use NLO results; the
stability of the perturbative expansion can be tested; the dependence
on the heavy masses of the theory can be precisely taken into account
and enforced. Moreover, the results obtained in the calculation of the
$\Delta S\!=\!1$ Effective Hamiltonian can be easily generalized to
the case of other effective hamiltonians, in the Standard Model or
even beyond it. 

For all these reasons and many others I do not detail
here, we think that this calculation will be important, even if,
as we will show in the phenomenological analysis, for
the time being the theoretical prediction for
$\epsi$ is not adequate to be competitive with the
improvements of the experimental measurements in the near future.

\section{The $\Delta S\!=\!1$ Effective Hamiltonian: theoretical 
construction}

I would like to report here about some main points in the theoretical
calculation of the effective hamiltonian which describes $\Delta
S\!=\!1$ decays and $\epsi$. I will recall the
general construction of the effective hamiltonian for the physical
problem at hand, the importance of the matching between
``short-distance'' and ``long-distance'' physics, the independence of
any physical result from the matching scale and its consequences. 
I will finally present the NLO solution of the evolution equation for
the Wilson coefficients and discuss the regularization scheme
independence of the final result. Full details and references about each 
single point can be found in \cite{cfmr2}. 

The most general effective hamiltonian can be written, using the
Operator Product Expansion (OPE) of the product of the two original
weak currents as:
\bea 
\!\!\!\!\!& &\langle F \vert {\cal H}_{eff}^{\Delta S\!=\!1}\vert I \rangle 
\propto\\
\!\!\!\!\!\! & &  \int d^4 x D_W(x^2, M_W^2) \langle F \vert T \left(
J_{\mu}(x),J^{\dagger}_{\mu}(0) \right)\vert I \rangle \nn \\
&\rightarrow& \sum_{i}\langle F \vert O_{ i}(\mu)  \vert I \rangle 
C_{ i}(\mu)  = \vec O^T(\mu) \vec C(\mu)\nn
\label{ope}
\eea
where $\{O_i(\mu)\}$ is a complete basis of independent operators
(which mix under renormalization, when QCD and QED corrections are
taken into account), appropriate to the physical problem at hand,
and $\{C_i(\mu)\}$ the related Wilson coefficients.
For the specific case of the $\Delta S\!\!=\!\! 1$ effective
hamiltonian, we have used the following basis made of vertex-type and
penguin-type operators:
\bea
O_{ 1}&=&({\bar s}_{\alpha}u_{\beta})_{\sss (V-A)}
    ({\bar u}_{\beta}d_{\alpha})_{\sss (V-A)}
   \nn\\
O_{ 2}&=&({\bar s}_{\alpha}u_{\alpha})_{\sss (V-A)}
    ({\bar u}_{\beta}d_{\beta})_{\sss (V-A)}
\nn \\
O_{ 3,5} &=& ({\bar s}_{\alpha}d_{\alpha})_{\sss (V-A)}
    \sum_{q=u,d,s,\cdots}({\bar q}_{\beta}q_{\beta})_{\sss (V\mp A)}
\nn \\
O_{ 4,6} &=& ({\bar s}_{\alpha}d_{\beta})_{\sss (V-A)}
    \sum_{q=u,d,s,\cdots}({\bar q}_{\beta}q_{\alpha})_{\sss (V\mp A)}
\nn \\
O_{ 7,9} &=& \frac{3}{2}({\bar s}_{\alpha}d_{\alpha})_{\sss (V-A)}
    \sum_{q=u,d,s,\cdots}e_{ q}({\bar q}_{\beta}q_{\beta})_{\sss (V\pm A)}
\nn \\
O_{ 8,10} &=& \frac{3}{2}({\bar s}_{\alpha}d_{\beta})_{\sss (V-A)}
    \sum_{q=u,d,s,\cdots}e_{ q}({\bar q}_{\beta}q_{\alpha})_{\sss (V\pm A)} 
    \nn \\
O^c_{ 1}&=&({\bar s}_{\alpha}c_{\beta})_{\sss (V-A)}
    ({\bar c}_{\beta}d_{\alpha})_{\sss (V-A)}
\nn \\
O^c_{ 2}&=&({\bar s}_{\alpha}c_{\alpha})_{\sss (V-A)}
    ({\bar c}_{\beta}d_{\beta})_{\sss (V-A)}
\label{opbasis}
\eea
where the subscript $(V \pm A)$ indicates the chiral structure and
$\alpha$ and $\beta$ are colour indices. The sum is intended over
those flavours which are active at the scale $\mu$.

The factorization of $\langle {\cal H}_{eff}^{\Delta S\!=\!1}\rangle$
in operator matrix elements and Wilson coefficients is completely
arbitrary. It depends on the scale $\mu$ at which we decide to match
the ``short-distance'' and ``long-distance'' physics.  But the final
physical amplitude must be $\mu$-independent. Imposing this condition,
a Renormalization Group Equation (RGE) for the Wilson coefficients can
be derived. When both QCD ($\alphas$) and QED ($\alphae$) radiative
corrections are taken into account, the RGE for the coefficients is:
\bea
\Bigl( - \frac {\partial} {\partial t} + \beta (\alphas)
\frac {\partial} {\partial \alphas} - \frac {\hat \gamma (\vec\alpha) }{2} 
\Bigr) \vec C(t, \alphas(t), \alphae)\!\!&=&\!\!0 \nn\\
& &
\label{rge} 
\eea
where (in the case of a weak interaction effective theory) $t=ln
(M_W^2 / \mu^2 )$, $\vec\alpha=(\alphas,\alphae)$ and in particular we
ignore the running of $\alphae$.  The solution of eq.(\ref{rge}) gives
the evolution of the coefficients with the mass scale $\mu$ and
depends on $\beta(\alphas)$ (the QCD $\beta$-function),
$\hat\gamma(\alphas,\alphae)$ (the QCD+QED anomalous dimension matrix)
and the value of the coefficients at a given initial scale (initial
conditions for the solution of the RGE).  All these quantities are
perturbative objects, expanded in powers of $\alphas$ and $\alphae$
(or in powers of $\alphas^k \alphae^h(\alphas t)^n(\alphae t)^m$ in a
Leading Logarithm expansion). Thus the solution of eq.(\ref{rge}) has
also to be specified at a given order, f.i. LO (Leading Order), NLO
(Next-to-Leading Order), etc.  Our result concerns the NLO evolution
of the Wilson coefficients.  Assuming the initial value of the
coefficients to be $\vec C(\Mw)$
\cite{flynn,anat1,cfmr2}, their value at the scale $\mu$ will be given
in term of the evolution matrix $\W[\mu,\Mw]$ as:
\be 
\vec C(\mu) = \W[\mu,\Mw] \vec C(\Mw) 
\label{evo} 
\ee
Consider the perturbative expansion of $\beta(\alphas)$ and
$\hat\gamma(\alphas,\alphae)$:
\be 
\beta(\alphas) = - \beta_0 \frac {\alphas^2} {4\pi}
- \beta_1 \frac {\alphas^3} {(4\pi)^2} + O(\alphas^4) 
\label{betafun}
\ee
\bea
\hat \gamma \!&=&\! \frac {\alphas }{ 4 \pi } \hat \gamma_s^{(0)} +
 \frac {\alphae }{4 \pi} \hat \gamma_e^{(0)}+ \nn\\
& & (\frac {\alphas }{4 \pi})^2 \hat \gamma_s^{(1)} +
 \frac{ \alphas }{4 \pi} \frac{ \alphae}{4 \pi}  \hat \gamma_e^{(1)}
\label{dimanom}
\eea
where $\hat \gamma_{s,e}^{(0,1)}$ are $(10\times 10)$ matrices
expressing the mixing under renormalization (at one and two loops
respectively) among the operators in eq.(\ref{opbasis}). Then
at the NLO in QCD and QED, when only one power of $\alphae$ is taken
into account ($\alphae^n$ negligible for $n>1$.  This means that terms
of order $(\alphas t)^n$, $\alphas(\alphas t)^n$, $(\alphae t)(\alphas
t)^n$ and $\alphae(\alphas t)^n$ are resummed in the Leading-Log
expansion), the evolution matrix $\W[\mu,\Mw]$ is of the form:
\be
\W[\mu,\Mw]= \M[\mu] \U[\mu,\Mw] \Mp[\Mw]
\label{monster} 
\ee
with:
\bea  
\M[\mu] &=& \left(\hat 1 +\frac{\alphae }{4\pi}\Ke\right)
  \left(\hat 1 +\frac{\alphas (\mu)}{4\pi}\J\right) \nn\\
& & \left(\hat 1+\frac{\alphae}{\alphas (\mu)}\PP\right)
\label{mo1} 
\eea
and
\bea 
\Mp[\Mw] &=& \left(\hat 1-\frac{\alphae}{\alphas (\Mw)}\PP\right) \\
  & & \left(\hat 1 -\frac{\alphas (\Mw)}{4\pi}\J\right) 
 \left(\hat 1 -\frac{\alphae }{4\pi}\Ke\right) \nn
\label{mo2} 
\eea
Here $\K$, $\J$ and $\PP$ are recursively determined by a set of
complicate equations I do not report here (see ref.
\cite{cfmr1,cfmr2} for more details) and results to be
functions of $\hat\gamma_{s,e}^{(0)}$, $\hat \gamma_{s,e}^{(1)}$,
$\beta_0$ and $\beta_1$. 

On the other hand, the NLO initial conditions $\vec C(\Mw)$, defined by
matching the full theory with the effective one at the $\Mw$ scale and
at order $O(\alphas,\alphae)$, are of the form:
\bea
\vec C(\Mw) \!\!&=&\!\! \vec T^{(0)} + \frac {\alphas} { 4 \pi}
\Bigl( \vec T^{(1)} - \hat r^T \vec T^{(0)} \Bigr)  + \nn\\
& & \frac {\alphae} { 4 \pi}
\Bigl( \vec D^{(1)} - \hat s^T \vec T^{(0)} \Bigr) 
\label{coeic} 
\eea
where $\vec T^{(1)}$ and $\vec D^{(1)}$ represent the coefficients
obtained calculating the one loop matching at the scale $\Mw$, while
$\hat r$ and $\hat s$ express the mixing in the operator matrix
elements, at the $\Mw$ scale and at the given order
($O(\alphas,\alphae)$). Both $\vec T^{(1)}$, $\vec D^{(1)}$, $\hat r$
and $\hat s$ depend on the external states chosen for the Feynman
diagrams involved in their calculation. However in the final
expression for the $\vec C(\Mw)$ coefficients this dependence cancels.
Any ``unbalanced'' modifications of the initial condition $\vec
C(\Mw)$ will break this cancellation and the dependence on the
external states must be recovered by the operator matrix elements at
the $\Mw$ scale, as we have indeed verified.

Our contribution has been the calculation of $\hat\gamma_{s,e}^{(1)}$
with respect to the operator basis in eq.(
\ref{opbasis}) and the development of the whole formalism summarized 
up to here in order to get the NLO answer for the Wilson coefficients.

We have worked in the modified Minimal Subtraction renormalization scheme
($\overline{MS}$) and in two different regularization schemes: Naive
Dimensional Reduction (NDR) and 't Hooft-Veltman scheme (HV). The
regularization scheme independence of the final result has been
demonstrated and explicitely checked. On this point we agree with the
only other similar calculation in the literature \cite{bjlw2,bjl},
while a difference on two single diagrams is still present (see
\cite{cfmr2} for further details). The scheme-dependent objects in the
calculation are those involving divergent parts of two-loop Feynman
diagrams or finite parts of one-loop Feynman diagrams, i.e. 
$\hat \gamma_{s,e}^{(1)}$ on one side and $\vec T^{(1)}$, $\vec
D^{(1)}$, $\hat r$ and $\hat s$ on the other (for the list of the
Feynman diagrams contributing at different levels see ref.
\cite{cfmr2}). However, they enter the final expression for the
evoluted coefficients $\vec C(\mu)$ in a scheme independent
combination.

The final results for the NLO coefficients can be given in any
definite regularization scheme, provided that the relative operator
matrix elements are also calculated in the same scheme, as is affordable
using Lattice QCD matrix elements. One could also use scheme
independent coefficients, but this would correspond to what we have
called an ``unbalanced'' definition of the initial conditions $\vec
C(\Mw)$, i.e. to an additional dependence of $\vec C(\mu)$ on the
external states, to be compensated by using matrix elements calculated
between the same external states.

\section{Results for $\epsi$}

Using the $\Delta S\!=\!1$ effective hamiltonian we have computed, we
can calculate the NLO amplitudes for Kaon decays into two pions:
$A(K\rightarrow\pi\pi\,\,(I=0))$ and $A(K\rightarrow\pi\pi\,\,(I=2))$,
whose imaginary parts enter the expression for $\epsilon^\prime$:
\bea 
\epsilon^{\prime}&=&\frac{e^{ i\pi/4}}{\sqrt{2}}\frac{\omega}
{\mbox{Re}A_{ 0}}\left[\omega^{ -1}(\mbox{Im}A_{ 2})^{\prime}-\right.\nn\\
& & \left. (1-\Omega_{ IB})\,\mbox{Im}A_{ 0}
\right]
\label{epsilonprime}
\eea
where $ \omega=\mbox{Re}A_{ 2}/ \mbox{Re}A_{ 0}=0.045 $ and:
\be
\mbox{Im}A_{ 2}=(\mbox{Im}A_{ 2})^{\prime}+\Omega_{IB}\mbox{Im}A_{ 0}
\ee
$\Omega_{ IB}\!=\!0.25 \pm 0.10$ denotes the isospin breaking 
contribution and we assume $\mbox{Re}A_0\!=\!3.3\cdot 10^{-7}$.

\begin{figure}[htb]
\epsfxsize=2.7in
\epsffile{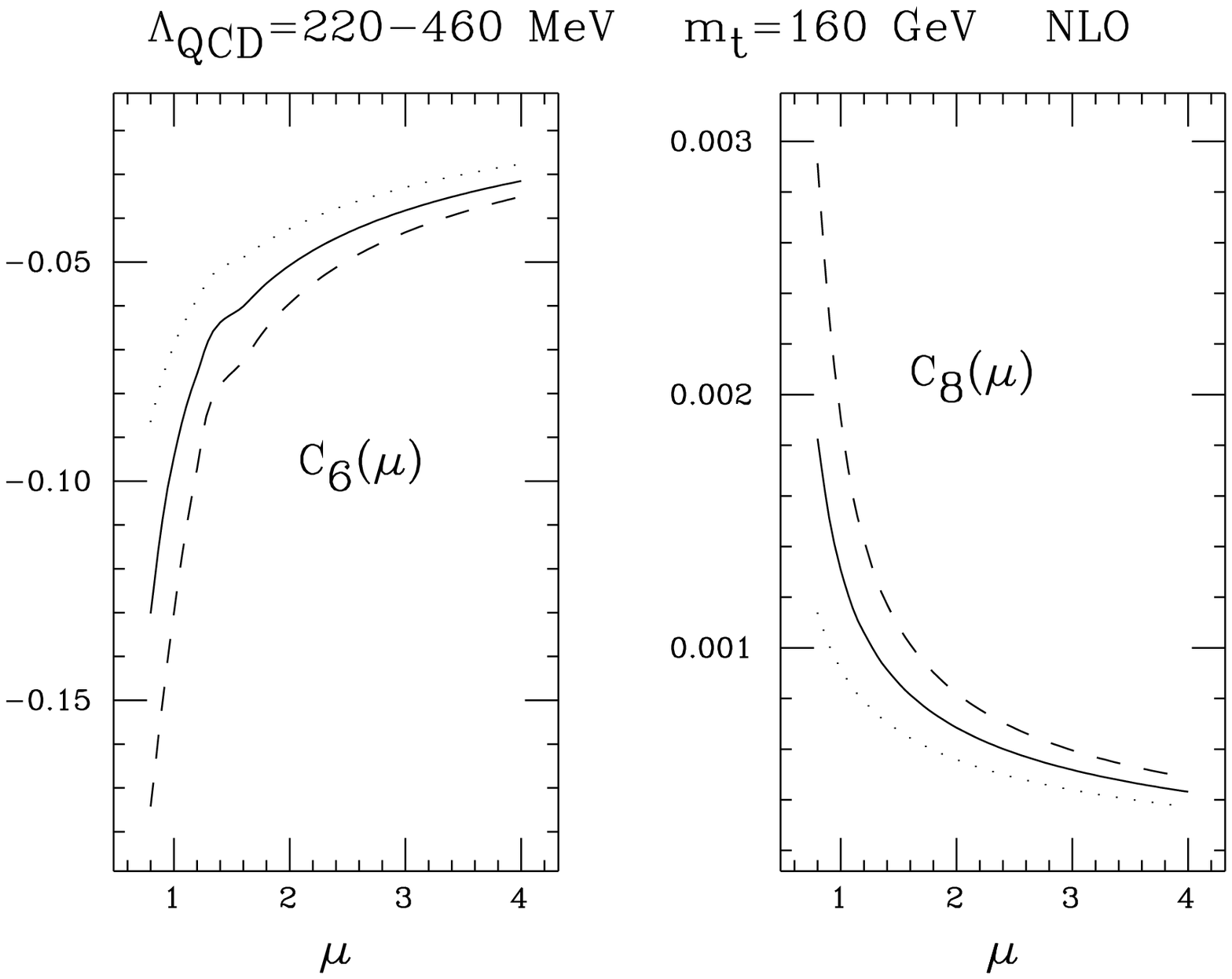}
\caption[]{ $C_6$ and $C_8$ as a function of $\mu$  for 
 $\Lambda_{QCD}=220$ (dotted),$ 340$ (solid) and $460$ (dashed) MeV.}
\label{ncf68}
\end{figure}

$\mbox{Im}A_{ 0}$ and $(\mbox{Im}A_{ 2})^{\prime}$ are expressed in
terms of Wilson coefficients and operator matrix elements.  We have
implemented a numerical program for the calculation of the evolution
of the Wilson coefficients at NLO and we have taken the matrix
elements of the operators in eq.(\ref{opbasis}) from Lattice QCD.  We
have chosen as matching scale $\mu\!=\!2$ GeV, because at this scale
the Wilson coefficients can be safely computed in a pure perturbative
regime and operator matrix elements, when calculable, can be taken
from Lattice QCD (this would not be true for other non-perturbative
methods, a much lower scale $\mu$ would be required, $\mu\sim 0.8-1.0$
GeV, where the Wilson coefficients show a quite strong instability --
see Fig.(\ref{ncf68})).

In particular, the matrix element of the generic operator $O_i$ is
given by:
\be
\langle O_i\rangle=B_i\langle O_i\rangle_{VIA}
\ee
where the subscript ``$_{VIA}$'' denotes the matrix element calculated
in the Vacuum Insertion Approximation and $B_i$ is the B-parameter of
the operator $O_i$, which quantitatively represents the deviation of
the physical matrix element from the VIA expectation. The VIA matrix
elements are calculated as functions of three quantities:
\bea
X\!&=&\!f_{\pi}\left(M_{ K}^{ 2}-M_{\pi}^{ 2}\right)\\
Y\!&=&\!f_{\pi}\left(\frac{M_{ K}^{ 2}}{\ms(\mu)+\md(\mu)}\right)^2\\
&\sim& 12\,X\left(\frac{0.15 \, \mbox{GeV}}{\ms(\mu)}\right)^2\nn\\
Z\!&=&\!4\left(\frac{f_{ K}}{f_{\pi}}-1\right)Y
\label{xyz}
\eea
while the B-parameters are in part taken from Lattice QCD (with a
given error) and, when this is not possible, they are allowed to vary
in a suitable range, consistent with some theoretical predictions or
experimental results. 

Thus, in the effective hamiltonian formalism the expressions for
$\mbox{Im}A_{ 0}$ and $(\mbox{Im}A_{ 2})^\prime$ result to be:
\bea
\label{ima0}
& &\!\!\!\! \mbox{Im}A_{ 0} =-\GF Im\Bigl({ V}_{ ts}^{ *}{ V}_{
td}\Bigr)\times \\
& &\!\!\!\!\left\{-\left(C_{ 6}B_{ 6}+\frac{1}{3}C_{ 5}B_{ 5}\right)Z
+\left(\phantom{\frac{1}{3}}C_{ 4}B_{ 4}+\right.\right. \nn\\
& &\!\!\!\!\left.\frac{1}{3}C_{ 3}B_{ 3}\right)X+C_{ 7}B_{ 7}^{ 1/2}
\left(\frac{2Y}{3}+\frac{Z}{6}-\frac{X}{2}\right)+ \nn\\
& &\!\!\!\!C_{ 8}B_{ 8}^{ 1/2}\left(2Y+\frac{Z}{2}+
\frac{X}{6}\right)-C_{ 9}B_{ 9}^{ 1/2}\frac{X}{3}+\nn\\
& &\!\!\!\!\left.\left(\frac{C_{ 1}
B_{ 1}^{ c}}{3}+C_{ 2}B_{ 2}^{ c}\right)X\right\}\nn
\eea
and
\bea
\label{ima2}
& &\!\!\!\!(\mbox{Im}A_{ 2})^{\prime}=-G_{ F}Im\Bigl({ V}_{ ts}^{ *}
{V}_{ td}\Bigr)\times \nn\\
& &\!\!\!\!\left\{C_{ 7}B_{ 7}^{ 3/2}\left(\frac{Y}{3}-\frac{X}{2}\right)+
\right.  \\
& &\!\!\!\!\left.C_{ 8}B_{ 8}^{ 3/2}\left(Y-\frac{X}{6}\right)+
C_{ 9}B_{ 9}^{ 3/2}\frac{2X}{3}\right\}\nn
\eea

As one can see from eqs.(17) and (18), $\epsi$ depends
on the analytical calculation of the coefficients (as functions of the
scale $\mu$, the top mass $m_t$ and $\Lambda_{QCD}$) and on some
experimental and theoretical input parameters, in particular: the
B-parameters, the strange quark mass $\ms$, the CKM matrix elements.
Many other well known quantities enter the determination of $\epsi$,
but they are clearly not so relevant in the analysis of the
uncertainties which still affect our theoretical analysis.

I would like to separate two main arguments. The first one concerns the
discussion of $\epsi$ in the more general contest in which we can
understand the mechanism of CP violation in the Standard Model (or
even beyond it). The second one, on the other hand, is more
specifically related to the determination of $\epsi$ itself, on the
basis of the NLO calculation of the Wilson coefficients
\cite{cfmr1,anat2}. 

\begin{figure}[htb]
\epsfxsize=2.7in
\epsffile{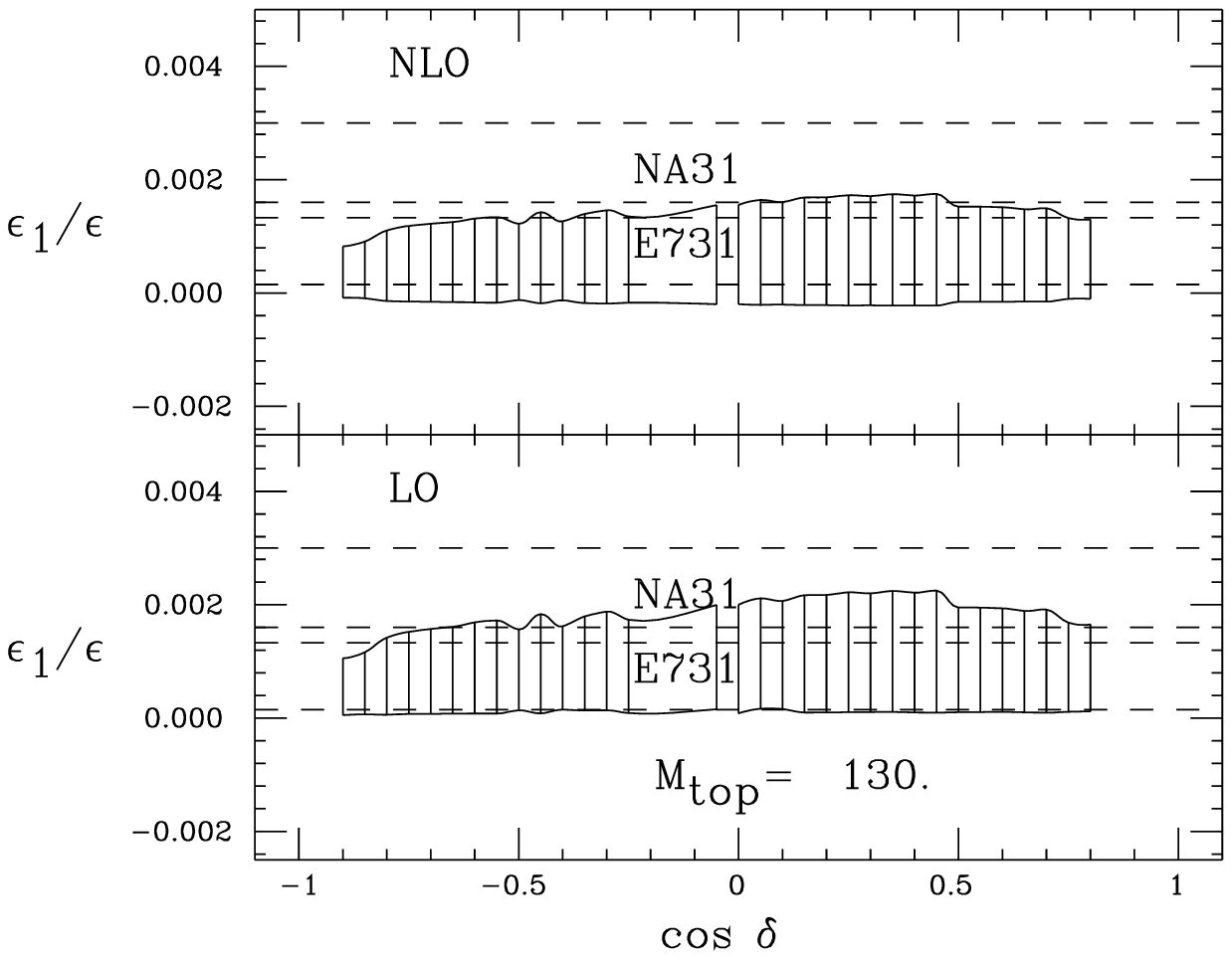}
\caption[]{LO and NLO predictions for $\epsi$:
        the band of the allowed values is shown , for $\mt=130$ GeV.
        The experimental values of NA31 and E731 are indicated (dashed lines).}
\label{mt130}
\end{figure}

\begin{figure}[htb]
\epsfxsize=2.7in
\epsffile{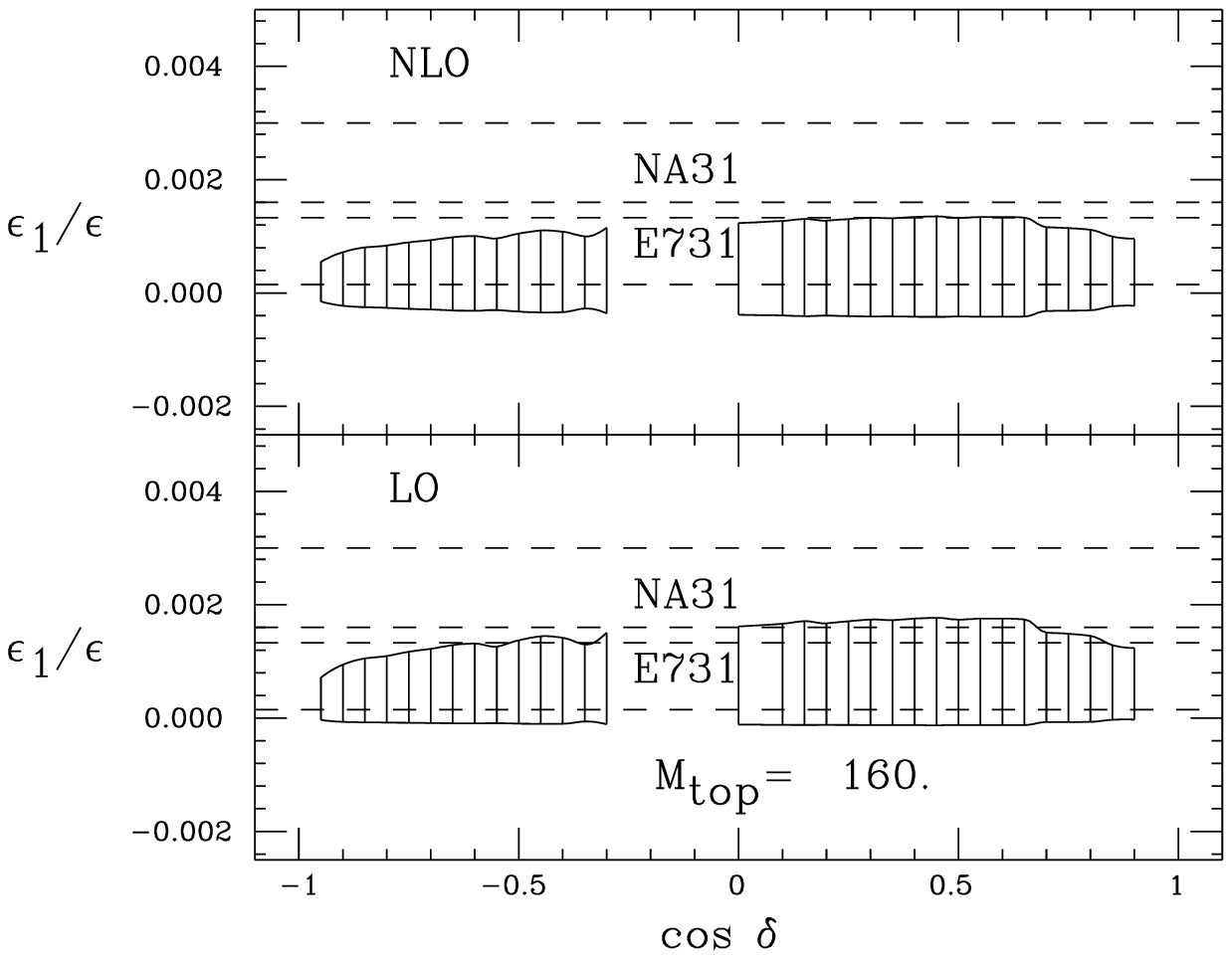}
\caption[]{LO and NLO predictions for $\epsi$:
        the band of the allowed values is shown , for $\mt=160$ GeV.
        The experimental values of NA31 and E731 are indicated (dashed lines).}
\label{mt160}
\end{figure}

\begin{figure}[htb]
\epsfxsize=2.7in
\epsffile{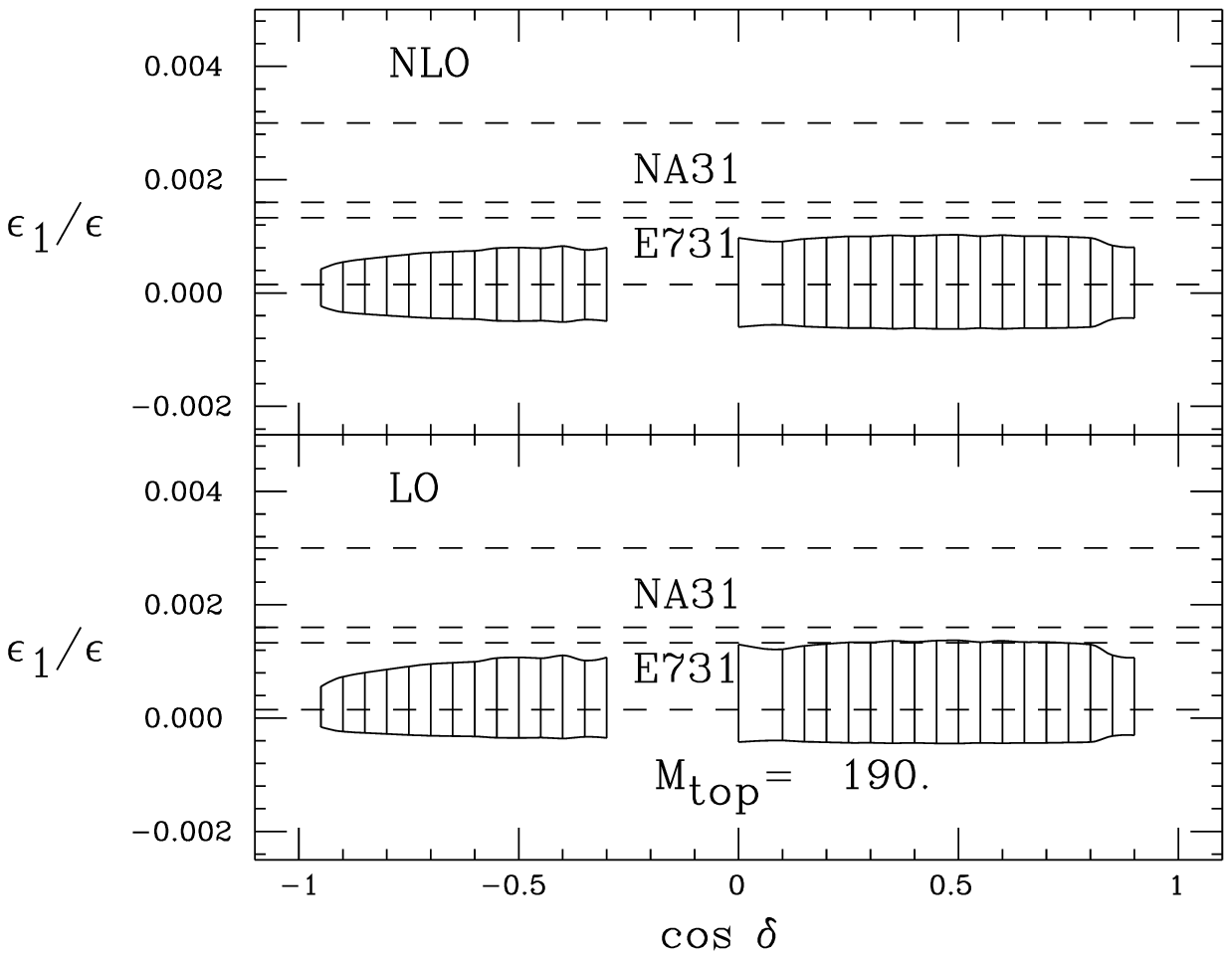}
\caption[]{LO and NLO predictions for $\epsi$:
        the band of the allowed values is shown , for $\mt=190$ GeV.
        The experimental values of NA31 and E731 are indicated (dashed lines).}
\label{mt190}
\end{figure}

In a global analysis of CP-violation within the Standard Model
framework, other physical quantities have to be taken into account,
mainly: the $\epsilon$ parameter of CP violation in the $K_0\!-\!\bar
K_0$ mixing and the $x_d$ parameter of $B_0\!-\!\bar B_0$ mixing
(related to $f_B$ or to the $B_B$ B-parameter). $\epsilon$ and $x_d$, 
as $\epsi$ depend on some CKM matrix elements and moreover
introduce a strong dependence on $B_K$ (the B-parameter for the
$K_0\!-\!\bar K_0$ mixing) and on the B-meson lifetime $\tau_B$. Many
phenomenological analysis of this kind exists in the literature, f.i.
\cite{anat1,topst,anat2,lmmr}. We are presently updating the analysis of
ref.\cite{lmmr}, both with respect to the values of the parameters
used and to the use of the NLO expression for $\epsi$ \cite{cfmr3}.
The main idea of ref.\cite{lmmr} is to use the common dependence of
$\epsilon$, $x_d$ and $\epsi$ on $\cos\delta$ ($\delta$ being the
CP-violating phase in the CKM matrix) in order to select between
positive and negative values of $cos\delta$. For a fixed value of $\mt$ 
(we have taken $\mt\!=\!(160\pm 30)$ GeV) the theoretical value of
$\epsilon$ -- obtained allowing $B_K$, $\Lambda_{QCD}$, $A$ and $\sigma$
(in the Wolfestein parametrization of the CKM matrix, see
f.i. \cite{lmmr}) varying in a $1\sigma$-interval around their central
values -- is fitted with the experimental one ($|\epsilon|\!=\!  2.268\cdot
10^{-3}$). For the selected values of $\cos\delta$, $f_B$ and $\epsi$
are calculated (we assume $x_d\!=\! 0.685\pm 0.0786$ and this time we
vary also $\tau_B$, $\Omega_{IB}$, $\ms$ and the B-parameters with the
previous criterion) and from the knowledge of both of them a single
region, $\cos\delta >0$ or $\cos\delta < 0$ should be selected. Indeed
the present proposed values for $f_B$:
\be
f_B=\left\{\brr{ll}
(205\pm 40) \mbox{MeV} & \mbox{\cite{abada}} \\
(195\pm 10\pm 30) \mbox{MeV} & \mbox{\cite{soni}} \\
(160\pm 6\pm \brr{l} +53 \\ -19 \err) \mbox{MeV} & \mbox{\cite{ukqcd}}
\err\right.
\ee
seem to select quite unambiguously the $\cos\delta >0$ region. On the
other and, as we will discuss more extensively in the following, this
is not the case for $\epsi$, at the present level of theoretical and
experimental uncertainty, see figs.(\ref{mt130})-(\ref{mt190}).

Let us add a few comments on the values used for some relevant
parameters. We use $\Lambda_{QCD}^{N_f=4}\!=\!(340\pm 120)$ MeV,
derived from \cite{alt} where the value for $\Lambda_{QCD}^{N_f=5}$
is given. 

We assume $\ms\!=\!(150\pm 30)$ MeV, lower than in ref.\cite{lmmr}, in
order to better compare our results to other recent analyses
\cite{anat2}, also if we still think that the status of uncertainty on
$\ms$ is such that the our ``old'' $\ms\!=\!(170\pm 30)$ MeV would
have been reasonably allowed.

Concerning the CKM matrix elements, the values of $A$ and $\sigma$
(Wolfestein par.) are derived respectively from $V_{cb}$ and
$\left|V_{ub}/V_{cb}\right|$. The value of $V_{cb}$ used is:
\be
V_{cb}= A\lambda^2= 0.043\pm 0.005
\ee
as extracted from incluse and exclusive semipletonic B-decays data
using a B-meson mean lifetime of:
\be
\tau_B= (1.5\pm 0.1)\cdot 10^{-12} \,\,\mbox{sec}
\ee
It is in agreement with refs. \cite{anat2,peccei1} and corresponds to:
\be
A=(0.9\pm 0.1)
\ee
The recent data presented at this conference
\cite{wither,tanaka,peccei2}, propose a little lower value of $V_{cb}$
($V_{cb}\!=\!(0.041\pm 0.005)$, corresponding to $A\!=\!(0.85\pm
0.10)$) and correspondentely a still higher value for the B-meson
mean lifetime $\tau_B$ ($\tau_B\!=\!(1.535\pm 0.025)$ ps). Finally,
our choice of $\left|V_{ub}/V_{cb}\right|$:
\be
\left|V_{ub}/V_{cb}\right| = \sigma\lambda= 0.085\pm 0.015
\ee
corresponds to a central value which mediates between the analysis of
the inclusive semileptonic B-spectrum based on the ACCMM or ISGW
models \cite{peccei1}. On the other hand, for the extimation of the
error we trust much more the ACCMM model. The corrisponding value of
$\sigma$ results to be:
\be
\sigma=0.39\pm 0.07
\ee
A much more detailed discussion will be given in ref.\cite{cfmr3}.

Let us now consider the NLO analysis of $\epsi$. The new values
assumed for the CKM parameters have clearly produced a change in the
central value of $\epsi$ (at different $\mt$) with respect to
ref.\cite{cfmr1}. The same holds also for the change in $\ms$.  For
the time being we have not yet ultimate the analysis of the specific
dependence on these parameters \cite{cfmr3}, but it will be clearly
important to have it in view of a more general discussion (see
previous argument). 

Then, $\epsi$ mainly depends on the values of the B-parameters and on
the NLO expression for the Wilson coefficients. The B-parameters are
certainly one of the major sources of uncertainty in the problem, being
related to the poor knowledge we still have of the long-distance hadronic
physics. A detailed discussion of the values used, see Table
\ref{bpar}, is given in refs.\cite{lmmr,cfmr1}. I am not going to
repeat it here, because no relevant improvement has been produced in
the meanwhile.
\begin{table*}[hbt]
\setlength{\tabcolsep}{1.5pc}
\newlength{\digitwidth} \settowidth{\digitwidth}{\rm 0}
\catcode`?=\active \def?{\kern\digitwidth}
\caption[]{Values of the $B$-parameters. Entries with a $^{ (*)}$
are reasonable extimates; the others are taken from Lattice QCD
calculations.}
\label{bpar}
\begin{tabular*}{\textwidth}{@{}l@{\extracolsep{\fill}}rrrrrr}
\hline\hline\\
$B_{ K},B_{ 9}^{ (3/2)}$ &  $B_{ 1-2}^{ c}$ &
 $B_{ 3,4}$ &
$B_{ 5,6}$ & $B_{ 7-8-9}^{ (1/2)}$ & $B_{ 7-8}^{ (3/2)}$
\\ \\\hline \\
$0.8\pm 0.2$ &  $0 - 0.15^{ (*)}$ & $1 -  6^{ (*)}$ &
$1.0\pm 0.2$ & $1^{ (*)}$ & $1.0\pm0.2$
\\ \hline\hline
\end{tabular*}
\end{table*}

On the other hand, considering the Wilson coefficients, we may focus
on two main points: their variation from LO to NLO and their
dependence (once the NLO expression is assumed) on some peculiar
quantities: $\mt$, $\Lambda_{QCD}$ and the matching scale $\mu$.
For the purpose of the following discussion, let us write $\epsi$ in 
the following form:
\be  
\epsi \sim R \times C_6 B_6 \Bigl(1-\sum_i \Omega_i\Bigr) 
\label{kfact} 
\ee
where the contribution of $O_6$ has been explicitely factorized.
Indeed, at LO, $\epsi$ results to be dominated by $O_6$, and we want
to verify if it is still the case at NLO, when electromagnetic
penguin operators are expected to play a very important role.  The
leading $\Omega$'s result always to be: $\Omega_2^c$, $\Omega_4$ and
$\Omega_{7,8,9}^{3/2}$. 

First, let us fix $\mt\!=\!160$ GeV, $\mu\!=\!2$ GeV,
$\Lambda_{QCD}^{(N_f=4)}\!=\!340$ MeV, and look at the effects of the
NLO evolution (QCD+QED). The only relevant variations appear for
$\Omega_{7,8}^{3/2}$: $\Omega_{7}^{3/2}$ varies from 0.40 (LO) to 0.66
(NLO) and $\Omega_{8}^{3/2}$ from -0.02 (LO) to -0.09 (NLO).  However,
the sum $\Omega_{7}^{3/2}+\Omega_{8}^{3/2}+\Omega_{9}^{3/2}$ does not
change in a sensible way from LO to NLO (from 0.20 to 0.30), while the
sum $\Omega_{2}^{c}+\Omega_{4}$ is constant and equal to 0.10. Thus,
although the contribution of each single electromagnetic penguin
operator is relevant, their global effect is not and $\epsi$ continues
to be dominated by the contribution of $O_6$ still at NLO level.
Being $C_6$ lowered by the NLO corrections here considered (see
fig.(\ref{ncf6})), also the central value of $\epsi$ is lowered going
from LO to NLO, as figs.(\ref{mt130})-(\ref{mt160}) show.
The value measured by the Fermilab collaboration at E731
\cite{lawr1,tsch} seems to be in better agreement with our results.
\begin{figure}[htb]
\epsfxsize=2.7in
\epsffile{fig6a_160.eps}
\caption[]{ $C_6$ as a function of $\mu$  for 
 $\Lambda_{QCD}\!=\!340$ and $\mt\!=\!160$ GeV at LO and NLO.}
\label{ncf6}
\end{figure}

Taking now the NLO values for the $C_i$ and looking to specific
dependences, we observe first of all a remarkable variation with
$\mt$, see again figs.(\ref{mt130})-(\ref{mt160}). Higher values of
$\mt$ correspond to lower values of $\epsi$. The present ``tendence''
of $\mt$ seems to point towards higher values instead then smaller
ones and this is another point in favour of a small value of $\epsi$.
Moreover, looking at fig.(\ref{ncf68}), we can see both the dependence
on $\Lambda_{QCD}$ and the dependence on $\mu$ for different values of
$\Lambda_{QCD}$. What is quite clear here is just the fact (already
stressed in justifying the necessity of a high matching scale for the
effective hamiltonian) that the coefficients really ``blow up'' for low
values of $\mu$, let us say below 1 GeV, indicating that the
perturbative approach is nomore reliable at that scale.

\section{Conclusions}

From the previous phenomenological analysis, it is clear that for the
time being the value of $\epsi$ is till compatible with zero and
nothing definitive can be stated neither experimentally nor
theoretically. Thus, the Standar Model prediction of a small but
definitely non-zero $\epsi$ could still be confirmed or not.  The
experimental scenario seems at the moment to be much more promising
\cite{tsch} than the theoretical one. What we have done put the
calculation of the perturbative part of the $\Delta S\!=\!1$ effective
hamiltonian on a more solid ground, while some sensible improvements
in the calculation of the long-distance physics of the problem is
mandatory.

\end{document}